\documentclass{llncs}
\usepackage{amssymb,amsmath,stmaryrd,mathrsfs}
\usepackage[ruled]{algorithm2e}
\usepackage{picins,hyperref}
\usepackage{mathtools}
\usepackage{subfigure,wrapfig,makeidx}
\usepackage{multirow}
\usepackage{tikz}
\usetikzlibrary{positioning}
\usetikzlibrary{chains}
\usetikzlibrary{decorations}
\usetikzlibrary{calc}
\usetikzlibrary{matrix}
\usetikzlibrary{scopes}
\usetikzlibrary{decorations.pathmorphing}
\usetikzlibrary{decorations.shapes}
\usetikzlibrary{shapes}
\usetikzlibrary{shapes.symbols}
\usetikzlibrary{decorations.pathreplacing}
\tikzstyle{base} = [>=stealth,thick]

\usepackage{setspace}
\AtBeginDocument{%
  \addtolength\abovedisplayskip{-0.5\baselineskip}%
  \addtolength\belowdisplayskip{-0.5\baselineskip}%
}
\usepackage{paralist}
\DeclareFontFamily{U}  {MnSymbolC}{}
\DeclareFontShape{U}{MnSymbolC}{m}{n}{
    <-6>  MnSymbolC5
   <6-7>  MnSymbolC6
   <7-8>  MnSymbolC7
   <8-9>  MnSymbolC8
   <9-10> MnSymbolC9
  <10-12> MnSymbolC10
  <12->   MnSymbolC12}{}
\DeclareSymbolFont{MnSyC}{U}{MnSymbolC}{m}{n}
\DeclareMathSymbol{\righthalfcup}{\mathrel}{MnSyC}{184}
\DeclareMathSymbol{\ostar}{\mathrel}{MnSyC}{102}

\newcommand{\Var}{\textsf{Var}}
\renewcommand{\phi}{\varphi}
\newcommand{\defeq}{\triangleq}

\newcommand{\IV}{\textsf{siv}}
\newcommand{\aff}{\textit{aff}}
\newcommand{\hull}{\textit{hull}}
\newcommand{\sem}[1]{\llbracket #1 \rrbracket}
\newcommand{\interp}[2]{\sem{#1}^{#2}}
\newcommand{\stable}{\mathit{stable}}

\newcommand{\entry}{v^{\textsf{entry}}}
\newcommand{\exit}{v^{\textsf{exit}}}
\newcommand{\mul}{\odot}
\newcommand{\plus}{\oplus}
\renewcommand{\star}{\ostar}

\newcommand{\tuple}[1]{\langle #1 \rangle}

\newcommand{\formula}[1]{\fbox{$#1$}}
\newcommand{\loopbody}{\phi_{\textsf{body}}}

\def\azadeh#1{}
\def\zak#1{}

\title{Compositional Invariant Generation via Linear Recurrence Analysis}
\author{Azadeh Farzan and Zachary Kincaid}
\institute{University of Toronto}
\begin{document}

\maketitle

\begin{abstract}
This paper presents a new method for automatically generating numerical
invariants for imperative programs. Given a program, our procedure computes a
binary input/output relation on program states which over-approximates the
behaviour of the program.  It is compositional in the sense that it operates
by decomposing the program into parts, computing an abstract meaning of each
part, and then composing the meanings.
Our method
for approximating loop behaviour is based on first approximating the meaning
of the loop body, extracting recurrence relations from that approximation, and then using the closed forms to approximate the loop.  Our
experiments demonstrate that on verification tasks, our method is competitive
with leading invariant generation and verification tools.
\end{abstract}

\section{Introduction} \label{sec:intro}

Compositional program analyses operate by decomposing a program into parts,
computing an abstract meaning of each part, and then composing the meanings.
Compositional analyses have a number of desirable properties, including
scalability, parallelizability, and applicability to incomplete programs.
However, compositionality comes with a price: since each program fragment is
analyzed independently of its context, the analysis cannot benefit from
contextual information.  This paper presents a compositional method for
numerical invariant generation which, despite loss of contextual information,
compares favourably with leading (non-compositional) verification techniques.


The analysis proposed in this paper aims to compute a \emph{transition
  relation} which over-approximates the behaviour of a given program.  The use
of transition relations in compositional analysis (e.g.,
\cite{Muller-Olm2004,Popeea2006,Ammarguellat1990,Monniaux2009,Kroening2008,Biallas2012})
stems from the fact that they can be composed: for example, consider a program
$P = P_1;P_2$ which consists of two sub-programs $P_1$ and $P_2$ which are
executed in sequence.  A transition invariant $\sem{P}$ for $P$ can be
computed by computing transition invariants $\sem{P_1}$ and $\sem{P_2}$ for
the subprograms and then taking $\sem{P}$ to be the relational composition:
$\sem{P} = \{ (s,s'') : \exists s'. (s,s') \in \sem{P_1} \land (s',s'') \in
\sem{P_2} \}.$

A crucial question is how to compute abstractions of loops (i.e., \emph{loop
  summaries} \cite{Kroening2008}).  Our analysis is based on a classical idea:
find recurrence relations for variables modified in the body of a loop, and
then use the closed forms for these recurrences as the abstraction of the loop.
The focus of research on recurrence analysis has mainly been on computing the
\emph{exact} behaviour of a (necessarily) limited class loops, e.g. loops
where the body is a sequence of affine assignments (see
Section~\ref{sec:relwork} for a discussion of related literature).  We shift
the goal to computing \emph{over-approximate} behaviour of \emph{arbitrary}
loops. The main novelty of our approach is to
make synergistic use of recurrence analysis and compositionality: on one hand,
recurrence analysis can be used to compute accurate transition formulas for
loops; on the other hand, transition formulas for loop \emph{bodies} can be
mined for recurrence relations to enable recurrence analysis.

Compositionality enables using recurrence analysis for arbitrary loops in two ways.  
First, the fact that the transition formula for a loop is computed from a transition formula for
its body makes the control structure of the loop irrelevant (e.g., whether it
is a sequence of assignments or contains branching or nested loops -- its
transition formula is just a formula).  Second, having access to a loop body formula when computing a loop summary opens the door to using
Satisfiability Modulo Theories (SMT) solvers to extract a broad range
\emph{semantic} recurrences.  In particular, our
analysis is able to exploit \emph{approximate recurrences} (inequations over
linear terms) to compute interesting loop invariants even for variables which
do not satisfy recurrence equations in the classical sense, thus extending
the applicability of recurrence-based invariant generation and overcoming a
major barrier in its practical use.

In summary, this paper presents a 
  compositional method for generating numerical invariants (polynomial
inequalities of unbounded degree among integer and rational variables) for
programs.  The main technical contributions are as follows.
\begin{compactenum}
\item We give a method for computing abstractions of loops using summaries for
  their bodies.  This allows our analysis to apply to arbitrary code
  (with nested loops, unstructured loops, and arbitrary branching).  It also
  makes it possible to use SMT solvers to extract \emph{semantic} recurrence
  relations rather than syntactic recurrences obtained by pattern-matching source code.
\item We identify a class recurrence (in)equations that can be efficiently
  extracted from loop bodies using SMT solving technology and solved using
  simple linear algebra.
\item We give a \emph{linearization} algorithm which enables tractable (but
  necessarily approximate) reasoning about non-linear formulas over rationals
  and integers (Section~\ref{sec:discussion}).
\item We collect ideas from a diverse range of sources (including algebraic
  program analysis \cite{Farzan2013}, recurrence analysis
  \cite{Ammarguellat1990,Kovacs2008,Ancourt2010}, linearization
  \cite{Mine2006}, and symbolic abstraction
  \cite{Reps2004,Monniaux2010,Li2014}), and synthesize them into a cohesive
  presentation which can be used as a foundation for futher research on
  recurrence analysis.
\end{compactenum}


We implemented linear recurrence analysis and used it to verify assertions for
a suite of benchmarks.  Linear
recurrence analysis is able to prove the correctness of more benchmarks in
this suite than any of the leading verification tools for integer programs.


\section{Overview} \label{sec:motivation}

\begin{figure}[t]
  \begin{center}
  \subfigure[Program text]{
    \begin{minipage}[b]{3.25cm}
      \scriptsize
    \texttt{r := x } \texttt{\color{red}// \textit{remainder}}\\
    \texttt{q := 0 } \texttt{\color{red}// \textit{quotient}}\\
    \texttt{while(r >= y):}\\
    \hspace*{0.5cm}\texttt{\color{red}// \textit{subtract y from r}}\\
    \hspace*{0.5cm}\texttt{t := y}\\
    \hspace*{0.5cm}\texttt{while(t != 0)}\\
    \hspace*{1cm}\texttt{r := r - 1}\\
    \hspace*{1cm}\texttt{t := t - 1}\\\\
    \hspace*{0.5cm}\texttt{q := q + 1}\\\\
    \texttt{assert(x = q*y + r)}
    \vspace*{0.25cm}
    \end{minipage}
  }
  \subfigure[Flow graph]{
  \begin{tikzpicture}[base,node distance=1.5cm]
  \scriptsize
  \node [rectangle,rounded corners,draw] (entry) {$\entry$};
  \node[rectangle,rounded corners,draw,right of=entry] (v1) {$v_1$};
  \node[rectangle,rounded corners,draw,right of=v1] (v2) {$v_2$};
  \node[rectangle,rounded corners,draw,below of=v2,xshift=-1.5cm] (v3) {$v_3$};
  \node[rectangle,rounded corners,draw,right of=v3] (v4) {$v_4$};
  \node[rectangle,rounded corners,draw,below of=v4,xshift=-1cm] (v5) {$v_5$};
  \node[rectangle,rounded corners,draw,right of=v5,xshift=0.5cm] (v6) {$v_6$};
  \node[rectangle,rounded corners,draw,right of=v4] (v7) {$v_7$};
  \node[rectangle,rounded corners,draw,right of=v2] (v8) {$v_8$};
  \node[rectangle,rounded corners,draw,right of=v8,xshift=1.25cm] (exit) {$\exit$};

  \path (entry) edge[->] node[above]{\texttt{r := x}} (v1);
  \path (v1) edge[->] node[above]{\texttt{q := 0}} (v2);
  \path (v2) edge[->] node[left]{\texttt{[r >= y]}} (v3);
  \path (v3) edge[->] node[below]{\texttt{t := y}} (v4);
  \path (v4) edge[->] node[below]{\texttt{[t = 0]}} (v7);
  \path (v4) edge[->] node[left]{\texttt{[t != 0]}} (v5);
  \path (v5) edge[->] node[below]{\texttt{r := r - 1}} (v6);
  \path (v6) edge[->] node[right]{\texttt{t := t - 1}} (v4);
  \path (v7) edge[->] node[right]{\texttt{q := q + 1}} (v2);
  \path (v2) edge[->] node[above]{\texttt{[r < y]}} (v8);
  \path (v8) edge[->] node[above]{\texttt{assert(x = q*y+r)}} (exit);

  \end{tikzpicture}}

  \end{center}
  \caption{An integer division program, computing a quotient and
    remainder.  Statements of the form $[\psi]$ represent \emph{assumptions};
    i.e., statements which block if $\psi$ does not hold.} \label{fig:example}
\end{figure}
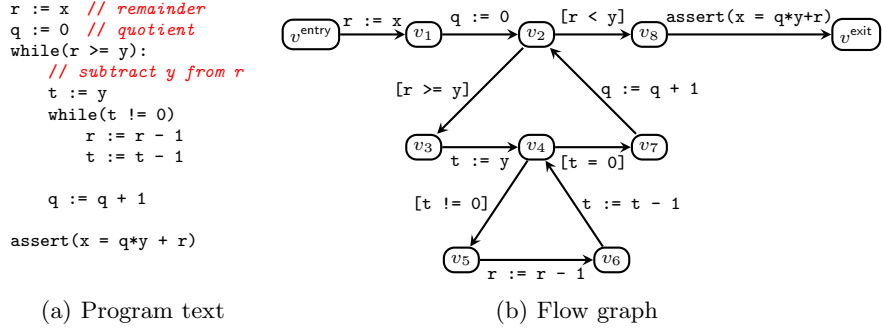

We will adopt a simple intraprocedural model in which a program is represented
by a control flow automaton (CFA) where edges are labeled by program
statements.  Figure~\ref{fig:example} depicts such a CFA for a program which
computes the quotient and remainder of division of a variable {\tt x} by a
variable {\tt y}.  We use this model for the sake of simplicity and to help
keep the presentation of our analysis short and self-contained.  We hope that
the basic idea behind the extension to procedures (implemented in the tool),
using the analysis to compute procedure summaries \cite{Sharir1981}, is clear without
formal explanation.



Our analysis, linear recurrence analysis (LRA), is presented in the algebraic
framework described in \cite{Farzan2013}.
Suppose that we wish to
prove that the assertion \texttt{assert(x = q*y + r)} always succeeds.  We begin by computing the set of paths from $\entry$ to $v_8$ (the location corresponding to the
{\tt assert} statement in the CFA).  This set of paths is represented 
by a \emph{path expression} for the vertex $v_8$, which is a regular expression over
an alphabet of control flow edges.  In principle, this can be accomplished by Kleene's
well-known algorithm for converting a finite automaton into a regular
expression \cite{Kleene1956} (but more efficient algorithms exist
\cite{Tarjan1981}).  For example, the following is a path
expression for $v_8$: {\small
\[\tuple{\entry,v_1}\!\cdot\!\tuple{v_1,v_2}\!\cdot\!\underbrace{\big(\tuple{v_2,v_3}\!\cdot\!\tuple{v_3,v_4}\!\cdot\!\overbrace{(\tuple{v_4,v_5}\!\cdot\!\tuple{v_5,v_6}\!\cdot\!\tuple{v_6,v_4})^*}^{\text{Inner loop}}\!\cdot \tuple{v_4,v_7}\!\cdot\!\tuple{v_7,v_2}\big)^*}_{\text{Outer loop}}\!\cdot\tuple{v_2,v_8}\]
}

\noindent Once we have a path expression representing the paths to $v_8$, we
compute an over-approximation of the executions to $v_8$ by {\em evaluating}
the path expression in some abstract domain.  The main benefit of this
algebraic framework is that an analysis is defined simply by providing an
interpretation for each of the regular expression operators (sequencing,
choice, and iteration, corresponding to the control structures of structured
programs), and then we may rely on a path expression algorithm
(\cite{Kleene1956,Tarjan1981}) to efficiently ``lift'' the analysis to
programs with arbitrary control flow.

Formally, a program analysis (in the framework
of \cite{Farzan2013}) is defined by an \emph{interpretation}, which consists
of a \emph{semantic algebra} and a \emph{semantic function}.  A semantic
algebra consists of a \emph{universe} which defines the space of possible
program meanings, and \emph{sequencing}, \emph{choice}, and \emph{iteration}
operators, which define how to compose program meanings.  A semantic function
is a mapping from control flow edges to elements of the universe which defines
the meaning of each control flow edge.
A path expression is \emph{evaluated} by interpreting the individual edges
using the semantic function, and interpreting the regular expression operators
using the corresponding operators of the semantic algebra (to compose
the interpretations of individual edges into interpretations of sets of program
paths).

Keeping this overall algorithm in mind, we proceed to describe the
interpretation which defines linear recurrence analysis.

\paragraph{\bfseries LRA Universe.}
The semantic universe of LRA (i.e., the space of program meanings) is the set
of (not necessarily linear) arithmetic \emph{transition formulas}.  If we let $\Var$ denote the set of
program variables and $\Var'$ the set of ``primed'' copies of program
variables, then a transition formula is an arithmetic formula with free
variables in $\Var \cup \Var'$.  Such a formula represents an input/output
relation between program states.

\paragraph{\bfseries LRA Semantic Function.}
The semantic function $\sem{\cdot}$ is a function that maps each edge of a
control flow automaton to its interpretation as a transition formula.  For
example (again, considering Figure~\ref{fig:example}), we have
\begin{align*}
  \sem{\tuple{\entry,v_1}} &= \formula{r' = x \land \stable(\{q, t, x, y\})}\\
  \sem{\tuple{v_1,v_2}} &= \formula{q' = 0 \land \stable(\{r, t, x, y\})}\\
  \sem{\tuple{v_2,v_3}} &= \formula{r > y \land \stable(\{q, r, t, x, y\})}
\end{align*}
where for $X \subseteq \Var$, we have $\stable(X) \defeq \formula{\bigwedge_{x
    \in X} x' = x}$; we use this to factor out equalities from the formulas
and make them more legible.  Boxes around formulas have no meaning, and are used only to make it easier to distinguish between equalities in formulas and the meta-language.

\paragraph{\bfseries LRA Operators.} The sequencing and choice operators of our analysis are defined as follows:
\begin{align*}
  \varphi \mul \psi &= \exists x''. \varphi[x''/x'] \land \psi[x''/x] &
  \text{Sequencing}\\ \varphi \plus \psi &= \varphi \lor \psi & \text{Choice}
\end{align*}
(where $\varphi[x''/x']$ denotes $\varphi$ with each primed variable $x'$
replaced by its double-primed counterpart $x''$, and $\psi[x''/x]$ similarly
replaces unprimed variables with double-primed variables).

The semantic function, sequencing, and choice operators are sufficient to
analyze loop-free code.  For example, we may consider how LRA
computes a transition invariant for the body of the inner loop of
Figure~\ref{fig:example}:
\begin{align*}
  \sem{\tuple{v_4,v_5}\cdot \tuple{v_5,v_6}} &= \sem{\tuple{v_4,v_5}} \mul \sem{\tuple{v_5,v_6}}\\
  &=\formula{t > 0 \land r'=r-1 \land \stable(\{q, t, x, y\})}
\end{align*}
\begin{align*}
  \sem{\tuple{v_4,v_5} \cdot \tuple{v_5,v_6} \cdot \tuple{v_6,v_4}} &=
  \sem{\tuple{v_4,v_5} \!\cdot\! \tuple{v_5,v_6}} \mul \sem{\tuple{v_6,v_4}}\\
  &= \formula{t > 0 \land r'=r-1 \land t'=t-1 \land \stable(\{q, x, y\})}
\end{align*}

The final step in describing our analysis is to provide a definition of the
iteration operator ($\star$) of LRA.  The idea behind the definition of the 
iteration operator is to use an SMT solver to extract recurrence relations
from the loop body, and then use the closed form of these recurrences for 
the abstraction of the loop. We explain this in detail in Section~\ref{sec:lra}.  Here, we illustrate how LRA works on the running example to provide some intuition
on the analysis.

\vspace*{-4pt}

\parpic[r]{
\begin{tabular}{|l|l|}
  \hline
  Recurrence & Closed form\\\hline
  $r' = r - 1$ & $r^{(k)} = r^{(0)} - k$\\
  $t' = t - 1$ & $t^{(k)} = t^{(0)} - k$\\ \hline
\end{tabular}
}

After computing a formula $\phi_{\textsf{inner}}$ representing the body of the
inner loop (as given above), we apply the iteration operator $\star$ to
compute a formula representing any number of executions of the inner loop.
The iteration operator begins by extracting the recurrence equations shown to
the right.  It then computes closed forms for these recurrences, also shown to
the right (where $x^{(k)}$ denotes the value that the variable $x$ takes on
the $k$th iteration of the loop). Note that this table omits ``uninteresting''
recurrences (such as $q' = q + 0$) which indicate that a variable does not
change in a loop.  These closed forms are used to abstract the loop as
follows:

\begin{align*}
  \varphi_{\textsf{inner}}^\star &=  \formula{\exists k. k \geq 0 \land r' = r - k \land t' = t - k \land \stable(\{q, x, y\})}\\
  &= \formula{r' = r + t' - t \land t' \leq t \land \stable(\{q, x, y\})}
\end{align*}


We may use this summary $\varphi_{\textsf{inner}}^\star$ for the inner loop to
compute a transition formula representing the body of the outer loop:
\begin{align*}
  \varphi_{\textsf{outer}} &= \sem{\tuple{v_2,v_3}} \mul \sem{\tuple{v_3,v_4}} \mul \varphi_{\textsf{inner}}^\star \mul \sem{\tuple{v_4,v_7}} \mul \sem{\tuple{v_7,v_2}}\\
  &= \formula{q' = q + 1 \land r' = r + t' - y \land t' = 0 \land r \geq y \land \stable(\{x, y\})} 
\end{align*}

\parpic[r]{
\begin{tabular}{|l|l|}
  \hline
  Recurrence & Closed form\\\hline
  $q' = q + 1$ & $q^{(k)} = q^{(0)} + k$\\
  $r' = r - y$ & $r^{(k)} = r^{(0)} - y^{(0)}k$\\ \hline
\end{tabular}
}
We then apply the iteration operator to compute a transition formula for the
outer loop.  The recurrences found for the outer loop and their closed forms
are shown to the right (again, with ``uninteresting'' recurrences omitted).
We note that our algorithm extracts these recurrences from
$\phi_{\textsf{outer}}$ using only semantic operations: the fact that
$\phi_{\textsf{outer}}$ is an abstraction of a looping computation is
completely transparent to the analysis.  Using the closed forms of the
recurrences to the right, we compute the following transition formula for the
outer loop:

\begin{align*}
  \varphi_{\textsf{outer}}^\star &=  \formula{\exists k. k \geq 0 \land q' = q + k \land r' = r - ky \land \stable(\{x, y\})}\\
  &= \formula{q' \geq q \land r' = r - (q'-q)y \land \stable(\{x, y\})}
\end{align*}


Finally, we compute a transition formula which approximates all executions
which end at $v_8$ as follows:
\begin{align*}
  \varphi_P &= \sem{\tuple{\entry,v_1}\cdot\tuple{v_1,v_2}} \mul \varphi_{\textsf{outer}}^\star \mul \sem{\tuple{v_2,v_8}}\\
  &= \formula{q' \geq 0 \land r' = x - q'y \land r \leq y \land \stable(\{x, y\})}
\end{align*}

This formula is strong enough to imply that assertion $x' = q'*y'+r'$ holds at
$v_8$.  This is particularly interesting because it requires proving a
\emph{non-linear} transition invariant for the loop, which is out of scope for
many state-of-the-art program analyzers.


\section{Abstracting Loops with Linear Recurrence Analysis} \label{sec:lra}

In this section, we describe the iteration operator of linear recurrence
analysis.  Suppose that we have a formula $\loopbody$ which approximates the
behaviour of the body of a loop.  Our goal is to compute a formula
$\loopbody^\star$ which represents the effect of zero or more executions of
the loop body.  Our iteration operator works by extracting recurrence
relations from the formula $\loopbody$ and then computing closed forms for
these relations. 
We present our iteration operator in three stages, based on the types of
recurrence relations being considered: \emph{simple recurrence equations},
\emph{stratified recurrence equations}, and \emph{linear recurrence
  (in)equations}.  Simple and stratified recurrences are classical classes of
recurrence equations.  Linear recurrence (in)equations generalize the class of
inequations presented in \cite{Ancourt2010} by using stratified recurrences to
generate polynomial (rather than just linear) inequations.  The main
conceptual contribution of this section is the idea to use SMT solvers to
extract recurrences (and other relevant information) from a loop body formula.

In the remainder of this section, we fix a formula $\loopbody$ representing
the body of a loop.  We assume that $\loopbody$ is expressed in linear
(rational and integer) arithmetic; our strategy for dealing with non-linear
arithmetic is described in Section \ref{sec:discussion}. We also assume that $\loopbody$ is satisfiable (if it is not, then we can take $\loopbody^\star$ to be $\bigwedge_{x \in \Var} x' = x$, which represents zero iterations of the loop).

\subsection{Simple recurrence equations}

We start by defining simple recurrences and induction variables.
\begin{definition}
A simple recurrence for a formula $\phi$ is an equation of the form $x' = x +
c$ (for a constant $c$) such that $\phi \models x' = x + c$.  If $x' = x + c$
is a simple recurrence for $\phi$, we say that $x$ \emph{satisfies} the
recurrence $x' = x + c$, and if there is some $c$ such that $x$ satisfies the recurrence $x' = x + c$, we say that $x$ is an \emph{induction variable}.
\end{definition}
Simple recurrences can be detected by first querying an SMT solver for
a model $m$ of $\loopbody$, and then asking whether $\loopbody$ implies $x' =
x + \interp{x'-x}{m}$ (where $\interp{x'-x}{m}$ denotes the interpretation of
the term $x'-x$ in the model $m$).  This implication holds iff $x$ is an
induction variable.

If $x$ is an induction variable that satisfies the recurrence $x' = x + c$,
then the closed form for $x$ is $x^{(k)} = x^{(0)} + kc$ (writing $x^{(k)}$ for
the value that $x$ obtains on the $k$th iteration of the loop).  To provide
some early intuition on the iteration operator to be developed in the
remainder of this section, let us suppose that we are only interested in
simple recurrences.  Then a possible definition for the iteration operator
is
\[ \loopbody^\star \defeq \formula{\exists k \geq 0. \bigwedge \{ x' = x + kc : x' = x + c \in \textit{SR}(\loopbody) \}} \]
where $\textit{SR}(\loopbody)$ is the set of simple recurrences satisfied by
$\loopbody$.

The iteration operator defined above is sound (it over-approximates the
behaviour of any number of iterations of the loop, since each variable is
either described exactly by a recurrence or is not constrained at all), but it
is imprecise.  The remainder of this section discusses more general
recurrence equations which can be used to compute more precise transition
invariants for loops.



\subsection{Stratified recurrences equations} \label{sec:stratified}

\parpic[fr]{\begin{minipage}{3cm}
    \noindent\textbf{while}\texttt{(x $\leq$ 10):}\\
    \noindent\hspace*{0.5cm}\texttt{x := x + 1}\\
    \noindent\hspace*{0.5cm}\texttt{y := y + x}\\
    \noindent\hspace*{0.5cm}\texttt{z := 2 * x}
    \vspace*{0.1cm}
  \end{minipage}}

Consider the loop shown to the right.  We can see that \texttt{x} satisfies a
simple recurrence equation $\texttt{x}' = \texttt{x} + 1$, and that \texttt{y}
satisfies a (non-simple) recurrence equation $\texttt{y}' = \texttt{y} +
\texttt{x} + 1$.  A closed form for \texttt{y}'s recurrence is $y^{(k)} =
\texttt{y}^{(0)} + \sum_{i=0}^{k-1} (\texttt{x}^{(i)} + 1)$.  Since \texttt{x}
satisfies a simple recurrence ($x' = x + 1$), we have a closed form for
$\texttt{x}^{(i)}$, so we may simplify this recurrence and remove the summation:

\[\texttt{y}^{(k)} =
\texttt{y}^{(0)} + \sum_{i=0}^{k-1} (\texttt{x}^{(0)} + i + 1) = \texttt{y}^{(0)} + k\texttt{x}^{(0)} + k + \sum_{i=0}^{k-1} i = \texttt{y}^{(0)} +
k\texttt{x}^{(0)} + \frac{k(k+1)}{2}.\]

\emph{Stratified recurrence equations} generalize this idea: starting from
simple recurrence equations, we solve more and more complicated recurrences
using the closed forms for simpler ones.  As with the example above,
stratified recurrences have non-linear closed forms.  Non-linear invariant
generation is not the main focus of our work, but it is sometimes a necessary
intermediate step for proving \emph{linear} invariants in a compositional
setting: since our analysis cannot take advantage of contextual information
when analyzing a loop, we generate a non-linear invariant and then, after the
analysis has examined more context, simplify it (using the linearization algorithm from
Section~\ref{sec:discussion}).

\begin{definition}
Let $\phi$ be a formula.  The stratified recurrence equations (and
stratified induction variables) of $\phi$ are defined inductively as:
\begin{compactitem}
\item A simple recurrence equation which is satisfied by $\phi$ is a
  stratified recurrence equation of $\phi$ (and a simple induction
  variable is a stratified induction variable) at stratum 0.
\item Let $\vec{y}$ denote a vector of the stratified induction variables of
  strata $\leq N$.  A recurrence of the form $x' = x + \vec{c}\vec{y}$ (where
  $\vec{c}$ is a vector of constants) is a stratified recurrence at stratum
  $N+1$ (and if $x$ satisfies such a recurrence, it is a stratified induction
  variable at stratum $N+1$).
\end{compactitem}
We use $\IV(\phi)$ to denote the set of all stratified induction
variables of $\phi$.
\end{definition}


\begin{algorithm}
  \small
  \SetKwInOut{Input}{Input}\SetKwInOut{Output}{Output}
  \SetKwComment{Comment}{/*}{*/}
  \Input{Satisfiable formula $\loopbody$}
  \Output{Affine hull of $\loopbody$}
  $H \gets \bot$;  $\psi \gets \loopbody$\;
  \While{there exists a model $m$ of $\psi$}{
    $H' \gets \bigwedge \{ x = \interp{x}{m} : x \in \Var \cup \Var' \}$\;
    $H \gets H \sqcup^= H'$ \Comment*[r]{Join in the domain of linear equalities}
    $\psi \gets \psi \land \lnot H$\;
  }
  \KwRet{$H$}
  \caption{Affine hull. \label{alg:affine_hull}}
\end{algorithm}

Let us now discuss how stratified recurrences are detected from a loop body formula $\loopbody$.  We begin by computing the affine hull
$\aff(\loopbody)$ of $\loopbody$
(Algorithm~\ref{alg:affine_hull}).\footnote{This algorithm is a specialization
  of the one in \cite{Reps2004} to the abstract domain of linear equalities.}

\begin{definition}
  The \emph{affine hull} $\aff(\phi)$ of a formula $\phi$ is the
  smallest affine set which contains $\phi$, represented as (the set of
  solutions to) a system of equations $A\vec{x} = \vec{b}$, where
  $\vec{x} = \begin{bmatrix*}x_1 & \dotsi & x_n & x_1' & \dotsi & x_n' \end{bmatrix*}$.
  Logically, $\aff(\phi)$ is a system of equations which satisfies
  the following three properties: (1) $\phi \models \aff(\phi)$, (2)
  every linear equation over $\Var \cup \Var'$ which is implied by $\phi$
  is also implied by $\aff(\phi)$, and (3) no equation in
  $\aff(\phi)$ is implied by the others.
\end{definition}



\zak{I think I can shorten this bit.  It would (possibly) make it a little less clear, but de-emphasize stratified recurrences.  Worth it?}
Our strategy for detecting stratified recurrences is based on the following
lemma.  Combined with property (2) of $\aff(\loopbody)$ above, this lemma
implies that any equation implied by $\loopbody$ can be expressed as a
linear combination of the equations in $\aff(\loopbody)$.

\begin{lemma}[\cite{Schrijver1986}, Corollary 3.1d]
  Let $A$ be a matrix, $\vec{b}$ be a column vector, $\vec{c}$ be a row
  vector, and $d$ be a constant.  Assume that the system $A\vec{x} = b$ has a
  solution.  Then $A\vec{x} = \vec{b}$ implies $\vec{c}\vec{x} = d$ iff there
  is a row vector $\vec{\lambda}$ such that $\vec{\lambda}A = \vec{c}$ and
  $\vec{\lambda}\vec{b} = d$.
\end{lemma}

Let us write $\aff(\loopbody)$ as $A\vec{x} = \vec{b}$.
Suppose that we have detected all recurrences of strata $< N$, and that we
want to determine whether a variable $x_i$ ($0 \leq i \leq n$) is an induction
variable at stratum $N$.  Then we ask whether there exists $\vec{\lambda}$,
$\vec{c}$, and $d$ such that:
\begin{compactitem}
\item $\vec{\lambda} A = \vec{c}$ and $\vec{\lambda}\vec{b} = d$ (i.e,
  $\vec{c}\vec{x} = d$ is implied by $\aff(\loopbody)$ and thus by $\loopbody$)
\item $c_i = 1$ and $c_{i+n} = -1$ (the coefficients of $x_i$ and $x_i'$ are 1 and -1, respectively)
\item For all $j$ such that $j \neq i+n$ and $n \leq j \leq 2n$, $c_{j} = 0$
  (except for $x_i'$, all coefficients of primed variables are 0).
\item For all $j$ such that $j \neq i$ such that $x_j$ is not an induction
  variable of strata $< N$ and $n \leq j \leq 2n$, $c_{j} = 0$ (except for
  $x_i$ and induction variables of strata $< N$, all coefficients for unprimed
  variables are 0).
\end{compactitem}
Thus, after computing the affine hull of $\loopbody$, determining whether a
given variable satisfies a stratified recurrence is simply a matter of solving
a system of linear equations (e.g., using Gaussian elimination).

\subsubsection{Closed forms for stratified recurrences.}  We first state a lemma:
\begin{lemma}
  The closed form for a stratified induction variable of strata $N$ is of the
  form \[x^{(k)} = p_0(k) + p_1(k)y_1^{(0)} + \dotsi + p_n(k)y_n^{(0)}\] where
  each $y_i$ is a stratified induction variable of strata $<N$ and each
  $p_i(k) \in \mathbb{Q}[k]$ is a polynomial of one variable with rational
  coefficients.
\end{lemma}

Our algorithm for solving stratified recurrences is based on a constructive
proof for this lemma.
We proceed by induction on strata.  The base case is trivial.  Suppose that we
have a recurrence at strata $N$ (and all $y_1,...,y_n$ are of strata $<N$):
$x' = x + c_1y_1 + \dotsi + c_ny_n + b$.  Then we may write
$x^{(k)} = x^{(0)} + \sum_{i=0}^{k-1} \big(c_1y_1^{(i)} + \dotsi + c_ny_n^{(i)} + b\big)$.
By our induction hypothesis, each $y_j^{(i)}$ can be written as a linear term
with coefficients from $\mathbb{Q}[k]$. It follows that there exists
$p_0,...,p_n \in \mathbb{Q}[k]$ so that 
\[ c_1y_1^{(i)} + \dotsi + c_ny_n^{(i)} + b = p_0(i) + p_1(i)y_{1}^{(0)} + \dotsi + p_n(i)y_n^{(0)}\]
Thus we have
\begin{align*}
  x^{(k)} &= x^{(0)} + \sum_{i=0}^{k-1} p_0(i) + p_1(i)y_{1}^{(0)} + \dotsi + p_n(i)y_n^{(0)}\\
  &= x^{(0)} + \sum_{i=0}^{k-1} p_0(i) + y_{1}^{(0)}\sum_{i=0}^{k-1} p_1(i) + \dotsi + y_{n}^{(0)}\sum_{i=0}^{k-1} p_n(i)
\end{align*}

The closed form of a summation of a polynomial of degree $m$ is a polynomial
of degree $m+1$.  We can find this polynomial via curve fitting (i.e., we
compute the first $m+1$ terms of the summation and then solve the
corresponding linear system of equations for the coefficients of the
polynomial).

\subsection{Linear recurrence (in)equations} \label{sec:lr_ineq}

\parpic[fr]{\begin{minipage}{4.5cm}
    \vspace{3pt}
    \noindent\hspace*{3pt}\textbf{while}\texttt{(x $\geq$ 0 $\land$ y $\geq$ 0):}\\
    \noindent\hspace*{0.5cm}\textbf{if}\texttt{(*):} \texttt{x := x - 1}\\
    \noindent\hspace*{0.5cm}\textbf{else}\texttt{:} \texttt{y := y - 1}
    \vspace{3pt}
  \end{minipage}}

Recurrence equations (such as the simple and stratified varieties) yield very
accurate approximations for \emph{some} variables, but what about variables
which do not satisfy \emph{any} recurrence equation?  For example, consider that neither \texttt{x} nor \texttt{y} satisfy a recurrence equation in
the loop to the right.  However, they \emph{do} satisfy recurrence
\emph{inequations}: $\texttt{x}-1 \leq \texttt{x}'$, $\texttt{x}' \leq
\texttt{x}$, $\texttt{y}-1 \leq \texttt{y}'$, and $\texttt{y}' \leq
\texttt{y}$.  These inequations can be closed to yield $\texttt{x}^{(0)} - k
\leq \texttt{x}^{(k)}$ and $\texttt{x}^{(k)} \leq \texttt{x}^{(0)}$, $y^{(0)}
- k\leq y^{(k)}$, and $y^{(k)} \leq y^{(0)}$.  In this section, we discuss
linear recurrence (in)equations, which allow us to compute good approximations
for loops that cannot be completely described by recurrence equations.

\begin{definition} \label{def:lri}
  A \emph{linear recurrence (in)equation} of a formula $\phi$ is an
  (in)equation which is implied by $\phi$ and which is of the form
  \[ \vec{c}\vec{x}' \bowtie \vec{c}\vec{x} + \vec{b}\vec{y} + d \]
  where $\mathop{\bowtie} \in \{ <, \leq, =\}$, $\vec{x}$ is any vector of
  variables, $\vec{y}$ is a vector of stratified induction variables in
  $\loopbody$, $\vec{c}$, $\vec{b}$ are constant vectors, and $d$ is a
  constant.
\end{definition}

Linear recurrence (in)equations generalize recurrence equations in two ways:
first, they allow for \emph{inequalities} rather than equations.  Second,
they allow recurrences for \emph{linear terms}, rather than just
variables.  For example, the linear recurrence equation $(\texttt{x}' +
\texttt{y}') = (\texttt{x} + \texttt{y}) + 1$ is satisfied by the body of the
loop above, which can be closed to yield $(\texttt{x}^{(k)} +
\texttt{y}^{(k)}) = (\texttt{x}^{(0)} + \texttt{y}^{(0)}) + k$.

We now describe our method for detecting and solving linear recurrence
(in)equations.  We begin by introducing a set of \emph{difference variables}
$\delta_x$, one for each variable $x \notin \IV(\loopbody)$ (variables which
do belong to $\IV(\loopbody)$ are already precisely described by recurrence
equations, so we need not approximate them).  We then compute (via
Algorithm~\ref{alg:convex_hull}) the convex hull of the formula $\psi$ defined
as:

\[\psi \defeq \exists X. \loopbody \land \bigwedge \{ \delta_x = x' - x : x \in \Var \setminus \IV(\loopbody) \} \]
where $X$ is $\Var' \cup (\Var \setminus \IV(\loopbody))$.

\begin{algorithm}
  \small
  \SetKwInOut{Input}{Input}\SetKwInOut{Output}{Output}
  \SetKwComment{Comment}{/*}{*/}
  \Input{Satisfiable formula $\psi$, set of variables $X$}
  \Output{Convex hull of $\exists X. \psi$}
  $P \gets \bot$\;
  \While{there exists a model $m$ of $\psi$}{
    Let $Q$ be a cube of the DNF of $\psi$ s.t. $m \models Q$\;
    $Q \gets \textit{project}(Q,X)$ \Comment*[r]{Polyhedral projection}
    $P \gets P \sqcup Q$ \Comment*[r]{Polyhedral join}
    $\psi \gets \psi \land \lnot P$\;
  }
  \KwRet{$P$}
  \caption{Convex hull. \label{alg:convex_hull}}
\end{algorithm}

Geometrically, the convex hull $\hull(\loopbody)$ is the smallest convex polyhedron
which contains $\loopbody$.  Logically, it is a set of (in)equations such that
(1) every (in)equation in $\hull(\loopbody)$ is implied by $\loopbody$, and (2) any
linear (in)equation (over $\Var \cup \Var'$) which is implied by $\loopbody$ is also implied
by $\hull(\loopbody)$.  For example, $\hull(\loopbody)$ for the loop above is:
\[ 0 \leq \delta_x \land \delta_x \leq 1 \land 0 \leq \delta_y \land \delta_y \leq 1 \land \delta_x + \delta_y = 1 \]

We note that the only variables which appear in the (in)equations in
$\hull(\loopbody)$ are (stratified) induction variables and difference variables.
Thus, we may write any (in)equation in $\hull(\loopbody)$ as $\vec{c}\vec{\delta}
\bowtie \vec{b}\vec{y} + d$ (where $\vec{\delta}$ is the vector of difference
variables, $\vec{y}$ is the vector of stratified induction variables,
$\vec{c}$ and $\vec{b}$ are constant vectors, and $d$ is a constant).
Recalling the definition of the difference variables, we may rewrite such an
inequation as $\vec{c}\vec{(\vec{x}'-\vec{x})} \bowtie \vec{b}\vec{y} + d$ and
then rewrite again as $\vec{c}\vec{x}' \bowtie \vec{c}\vec{x} + \vec{b}\vec{y}
+ d$, which matches the definition of linear recurrence (in)equations given
in Definition~\ref{def:lri}.

We may close such a linear recurrence (in)equation as follows:
\[\vec{c}\vec{x}^{(k)} \bowtie \vec{c}\vec{x}^{(0)} + \sum_{i=0}^{k-1} \vec{b}\vec{y}^{(i)} + d\]
We can compute a closed form for the summation $\sum_{i=0}^{k-1}
\vec{b}\vec{y}^{(i)} + d$ as in the preceding section.




\subsection{Loop guards}

A loop body typically contains crucial information about the execution of the
loop that cannot be captured by recurrence relations.  For example, consider
the loop in Section~\ref{sec:stratified}.  Supposing that the loop executes
$n$ times, we must have that $\texttt{x}^{(k)} \leq 10$ for each $k < n$.
Further, consider that the variable \texttt{z} is a function of the simple induction
variable \texttt{x}, and so $\texttt{z}^{(k)}$ can be described precisely in
terms of the pre-state variables (even though it does not itself satisfy any
recurrence):
\[ \texttt{z}^{(k)} = 
\begin{cases}
  \texttt{z}^{(0)} & \text{if } k = 0\\
  2(\texttt{x}^{(0)} + k + 1) & \text{otherwise.}
\end{cases} \]
The question is: how can we recover this type of information from a loop body
formula?

We define the \emph{guard} of a transition formula $\phi$ as follows:
\[guard(\phi) \defeq \formula{(\exists \Var. \phi) \land (\exists \Var'. \phi)}\]
If $\phi$ is a loop body formula, then $guard(\phi)$ is a formula which
over-approximates the effect of executing \emph{at least one} execution of the
loop.  Intuitively, $(\exists \Var. \phi)$ as a precondition that must hold
before every iteration of the loop and $(\exists \Var'. \phi)$ as a
post-condition of the loop that must hold after each iteration.

Consider again the example loop in Section~\ref{sec:stratified}, we have the following loop body formula
\[ \phi_{\textsf{body}} = \formula{x \leq 10 \land x' = x + 1 \land y' = y + x' \land z' = 2x'} \]
We compute $guard(\phi_{\textsf{body}})$ as follows:
\begin{align*}
  guard(\phi_{\textsf{body}}) &= \formula{(\exists \texttt{x},\texttt{y},\texttt{z}. \phi_{\textsf{body}}) \land (\exists \texttt{x'},\texttt{y'},\texttt{z'}. \phi_{\textsf{body}})} \\
  &\equiv \formula{(\texttt{x} \leq 10) \land (\texttt{x'} \leq 11 \land \texttt{z} = 2\texttt{x'})}\ ,
\end{align*}
and thereby recover the desired information about \texttt{x} and \texttt{z}.

Since loop body formulas may be large, it may be adventageous in practice to
simplify the guard formula by eliminating the quantifiers (as we did above).
A second option, which is more efficient but less precise, is to
over-approximate quantifier elimination.  Two possibilities are to use
Algorithm~\ref{alg:convex_hull} to compute the convex hull of
$guard(\loopbody)$, or to use optimization modulo theories \cite{Li2014} to
compute intervals for each pre- and post-state variable in
$\phi_{\textsf{body}}$.








\subsection{Bringing it all together}

We close this section by describing how the pieces defined in this section fit
into the iteration operator of linear recurrence analysis.  We let
$\textit{CR}(\loopbody)$ denote the set of closed linear recurrence
(in)equations (including simple and stratified recurrence equations) satisfied
by $\loopbody$.  Each such
(in)equation is of the form $\vec{c}\vec{x}^{(k)} \bowtie t$, where the free
variables of $t$ are drawn from $\{ x^{(0)} : x \in \Var \}$ and a
distinguished variable $k \notin \Var$ indicating the loop iteration.  We
define
\[\loopbody^+ \defeq \formula{\exists k. k \geq 1  \land \bigwedge \{ \vec{c}\vec{x'} \bowtie t[\vec{x}^{(0)} \mapsto \vec{x}] : \vec{c}\vec{x}' \bowtie t \in \textit{CR}(\loopbody) \}}\]
where $t[\vec{x}^{(0)} \mapsto \vec{x}]$ denotes the term $t$ with every
variable of the form $x^{(0)}$ is replaced by the corresponding variable $x$.

Finally, our iteration operator is defined as:
\[ \loopbody^\star \defeq \formula{(\loopbody^+ \land \textit{guard}(\loopbody)) \lor \bigwedge_{x \in \Var} x' = x.}\]


\section{Linearization} \label{sec:discussion}

The iteration operator presented in the previous section relies heavily on
using an SMT solver to extract information from loop body formulas.  This
strategy requires that loop body formulas are expressed in a decidable theory
which is supported by SMT solvers (in particular, linear arithmetic).
However, a program may contain non-linear instructions, and even if it does
not, our iteration operator may introduce non-linearity (consider
Example~\ref{fig:example}, where the transition formula for the outer loop
$\phi_{\textsf{outer}}^\star$ contains the non-linear proposition $r' =
x-q'y$).  Our solution to this problem is to \emph{linearize} non-linear
formulas before passing them to the iteration operator.

Linearization is an operation that, given an (arbitrary) arithmetic formula
$\phi$, computes a formula $\textit{lin}(\phi)$ which over-approximates $\phi$
(i.e., $\phi \Rightarrow \textit{lin}(\phi)$), but which is expressed in
linear arithmetic.  There is generally no best approximation of a non-linear
formula as a linear formula, so our method is (necessarily) a heuristic.

We explain our linearization algorithm informally using an example.  Consider the
following non-linear formula (where $w,x,y,z$ are integers):
\[ \psi \defeq 1 \leq w = x < y < 5 \land w * y \leq z \leq x * y \]
Our algorithm begins by normalizing $\psi$, separating it into a linear part
and a set of non-linear equations (introducing existentially quantified
temporary variables as necessary).  For example, the result of normalizing
$\psi$ is:
\[ \big(1 \leq w = x < y < 5 \land \leq \gamma_0 \leq z \leq \gamma_1 \big) \land \big( \gamma_0 = w*y \land \gamma_1 = x*y \big) \]

The left conjunct is a linear over-approximation of $\psi$, but it is very
imprecise: semantically equal (but syntactically distinct) non-linear terms become semantically \emph{un}equal in the over-approximation, and all information about the magnitude of non-linear terms is lost.
To increase precision of this approximation, we use two
strengthening steps.
\begin{compactenum}
\item We replace the non-linear operations with uninterpreted function symbols
  and then compute the affine hull of the resulting formula to infer
  equalities between non-linear terms.  For our example $\psi$, the we
  discover that $\gamma_0 = \gamma_1$.
\item We compute concrete and symbolic intervals for non-linear terms.
  Consider $\gamma_0 = x*y$ from our example $\psi$.  We first compute
  concrete ($x \in [1,3]$ and $y \in [2,4]$) and symbolic ($x \in [x,x]$ and
  $y \in [y,y]$) intervals for the operands $x$ and $y$, using symbolic
  optimization \cite{Li2014} to compute the concrete intervals.  We obtain a
  concrete interval for $x*y$ ($x*y \in [2,12]$) by multiplying the concrete
  intervals of its operands.  We obtain symbolic intervals for $x*y$ ($x*y \in
  [y,3y]$ and $x*y \in [2x, 4x]$) by multiplying the concrete interval for $x$
  by the symbolic interval for $y$ and vice-versa.  As a result of interval
  computation, we discover:
  $2 \leq \gamma_1 \leq 12 \land y \leq \gamma_1 \leq 3y \land 2x \leq \gamma_1 \leq 4x$
\end{compactenum}

\vspace*{1pt}
\noindent Finally, we take $\textit{lin}(\psi)$ to be the initial coarse linear
approximation of $\psi$ conjoined with the facts discovered by the two
strengthening steps.

We expect linearization to have broad applications outside of the context
in which we presented it, particularly in program analysis, where
over-approximation can be tolerated but non-linear terms cannot.  Finding
improved linearization heuristics is an interesting direction of future work.

\section{Experiments}\label{sec:experiments}

We wrote a tool which implements LRA and analyzes C code (using the CIL
\cite{cil} frontend).\footnote{The tool and benchmarks are available at \url{http://cs.toronto.edu/~zkincaid/lra}.}
We use Z3 \cite{z3} to resolve SMT queries that result
from applying the iteration operator and checking assertion violations.
Polyhedra operations are passed to the New Polka library implemented in Apron
\cite{apron}.  The quantifier elimination algorithm from \cite{Monniaux2010}
is used to compute loop guards.

We tested two different configurations of LRA: one which is fully
compositional ({\sc LRA-Comp}) and does not take advantage of contextual
information, and one ({\sc LRA}) which uses an intraprocedural polyhedron
analysis \cite{Cousot1978} to gain \emph{some} contextual information, but
which is otherwise compositional.  We compare LRA's performance against the
state-of-the-art invariant generation and verification tools 
{\sc CPAChecker} (overall winner of the 2015 Software Verification Competition) and {\sc SeaHorn} (winner of the loops category among tools which are sound for verification).

To evaluate the precision of LRA we used it to verify the correctness of a
suite of 119 small loop benchmarks of varying difficulty.  Our benchmark suite
was drawn from the \emph{loops} category of the 2015 Software Verification
Competition (SVComp-15), as well as a set of \emph{non-linear} benchmarks
(Non-linear), such as the one in Figure~\ref{fig:example}.  The results for
the 81 safe, integer-only benchmarks from these suites are shown in
Table~\ref{tab:lra}.  The suite also contains 38 \emph{unsafe} benchmarks: {\sc LRA} and {\sc LRA-Comp} have no false
negatives on these benchmarks; {\sc CPAChecker} has 3 and {\sc SeaHorn} has
2.

\begin{table}[!t]
\begin{center}
\begin{tabular}{|l|c||c|c|c|c|}
\hline
 Benchmark suite   & \# Bench  &  {\sc \scriptsize LRA} & {\sc\scriptsize LRA-Comp}   & {\sc \scriptsize CPAChecker} & {\sc \scriptsize SeaHorn} \\ \hline \hline

SVComp-15         &  74 &  65  & 60    & 37 & 65 \\
Non-linear & 7 & 6 & 5 & 1 & 3 \\\hline
Total & 81 & 71 (88\%)  & 65 (80\%) & 38 (47\%)  & 68 (85\%) \\
\hline
 \multicolumn{6}{c}{Running time across all benchmark suites}\\
 \hline
 \multicolumn{2}{|l|}{Mean} & 5.4s & 3.0s & 42.4s & 37.7s\\
 \multicolumn{2}{|l|}{Median} & 0.8s & 0.8s & 1.6s & 0.2s\\
 \hline
\end{tabular}
\end{center}
\caption{Experimental results.}
\label{tab:lra}
\end{table}


Our results demonstrate that LRA is an effective invariant generation
algorithm.  Even the fully compositional variant of LRA ({\sc LRA-Comp}) is
able to prove safety for 80\% of the benchmarks we considered).  We also note
that there are 8 benchmarks for which LRA can prove safety but which {\sc
  CPAChecker} and {\sc SeaHorn} cannot.


\section{Related work} \label{sec:relwork}

There is a great deal of work on compositional invariant generation and
acceleration which is related to the technique described in this paper.  In
this section, we compare our technique to a sampling of this work.

\noindent\emph{\textbf{Recurrence analysis.}}  The idea of using closed forms
of recurrence relations to approximate loops has appeared in a number of other
papers.  Generally speaking, our work differs from previous work in two
essential ways: first, we use an SMT solver to extract \emph{semantic}
recurrences, rather than \emph{syntactic} recurrences.  Second, we consider
\emph{approximate recurrences} (inequations over linear terms) rather than
exact recurrences (equations over variables).  A survey of some of this work
follows.

Ammarguellat and Harrison present a method for detecting induction variables
which is compositional in the sense that it uses closed forms for inner loops
in order to recognize nested recurrences \cite{Ammarguellat1990}.  Maps from
variables to symbolic terms (effectively a symbolic constant propagation
domain) is used as the abstract domain.  Kov\'{a}cs presents a technique for
discovering invariant polynomial equations based on solving recurrence
relations \cite{Kovacs2008}.  The simple and stratified recurrence equations
considered in this paper are a strict subset of the recurrences considered in
\cite{Kovacs2008}, but our algorithm for solving recurrences is simpler.
Kroening et al. \cite{Kroening2013} presents a technique for computing
\emph{under}-approximations of loops which uses polynomial curve-fitting to
directly compute closed forms for recurrences rather than extracting
recurrences and then solving them in a separate step.

Ancourt et al. present a method for computing recurrence inequations for while
loops with affine bodies \cite{Ancourt2010}.  Like the method we present on
Section~\ref{sec:lr_ineq}, their method is based on using difference variables
and polyhedral projections.  Our method generalizes this work by (1) extending
it to arbitary control flow, with (possibly non-linear) formulas as bodies
rather than affine transformations, (2) integrating recurrence inequations
with stratified induction variables, thereby allowing enabling the computation
of invariant polynomial inequations.  Ancourt et al. briefly discuss a
  method for computing invariant polynomial inequations, but it is based on
  higher-order differences rather than stratified recurrence inequations.  For
  example, in Figure~\ref{fig:example}, the analysis discussed in
  \cite{Ancourt2010} would be able to prove that \texttt{r} is decremented by
  a constant amount at every loop iteration, but could not prove that the
  constant amount is exactly \texttt{y}.

\noindent\emph{\textbf{Acceleration}.}  \emph{Acceleration} is a technique
closely related to recurrence analysis that was pioneered in infinite-state
model checking \cite{Boigelot1994,Finkel2002,Bardin2005}, and which has
recently found use in program analysis
\cite{Gonnord2006,Leroux2007,Jeannet2014}.  Given a set of reachable states
and an affine transformation describing the body of a loop, acceleration
computes an \emph{exact} post-image which describes the set of reachable
states after executing any number of iterations of the loop (although there is
recent work on \emph{abstract acceleration} uses computes over-approximate
post-images \cite{Gonnord2006,Jeannet2014}).  In contrast, our technique is
\emph{approximate} rather than exact, and computes loop summaries rather than
post-images.  A result of these two features is that our analysis to be
applied to arbitrary loops, while acceleration is classically limited to
simple loops where the body consists of a sequence of assignment statements.




\noindent\emph{\textbf{Compositional program analysis.}}  Compositional
program analysis has a long history.  Particular examples are interprocedural
analyses based on summarization \cite{Sharir1981} and elimination-style
dataflow analyses (a good overview of which can be found in \cite{Ryder1986}).
The following surveys recent work on compositional analysis for numerical
invariants.

Kroening et al. \cite{Kroening2008} and Biallas et al. \cite{Biallas2012}
present compositional analysis techniques based on predicate abstraction.
In addition to predicate abstraction, there are a few papers which use numerical abstract domains for compositional
analysis. These include an algorithm for detecting affine
equalities between program variables \cite{Muller-Olm2004}, an
algorithm for detecting polynomial equalities between program variables
\cite{Colon2004}, a disjunctive polyhedra analysis which uses
widening to compute loop summaries \cite{Popeea2006}, and a method for
automatically synthesizing transfer functions for template abstract domains
using quantifier elimination \cite{Monniaux2009}. Our abstract domain is the
set of arbitrary arithmetic formula, which is more expressive than these
domains, but which (as usual) incurs a price in performance.  It would be
interesting to apply abstractions to our formulas to improve the performance
of our analysis.



\noindent\emph{\textbf{Linearization.}}  Our linearization algorithm was
inspired by Min\'{e}'s procedure for approximating non-linear abstract
transformers \cite{Mine2006}.  Min\'{e}'s procedure abstracts non-linear terms
by linear terms with interval coefficients using the abstract value in the
pre-state to derive intervals for variables.  Our algorithm abstracts
non-linear terms by sets of symbolic and concrete intervals, and applies to the more
general setting of approximating arbitrary formulas.

\section{Conclusion} \label{sec:conclusion}

This paper presents a fully compositional algorithm for generating numerical
invariants of imperative programs.  Our method for abstracting loops makes
essential use of compositionality: we assume that we are given a formula which
approximates the body of a loop, and we use an SMT solver to extract
recurrence relations and then use the closed forms of these recurrences to
approximate the loop.  We have demonstrated experimentally that our method is
competitive with leading invariant generation and verification tools.


\end{document}